\documentclass[lettersize,journal]{IEEEtran}
\usepackage{amsmath,amsfonts}
\usepackage{array}
\usepackage[caption=false,font=normalsize,labelfont=sf,textfont=sf]{subfig}
\usepackage{textcomp}
\usepackage{stfloats}
\usepackage{url}
\usepackage{verbatim}
\usepackage{orcidlink}
\usepackage{graphicx}
\usepackage{algorithm}
\usepackage{algpseudocode}
\usepackage{amsfonts}
\usepackage{mwe} 
\usepackage{hyperref}
\usepackage{enumitem}
\usepackage{xcolor}   
\setlist{nosep, leftmargin=14pt}
\usepackage{amsfonts}
\usepackage{mwe} 
\usepackage{hyperref}

\def\BibTeX{{\rm B\kern-.05em{\sc i\kern-.025em b}\kern-.08em
    T\kern-.1667em\lower.7ex\hbox{E}\kern-.125emX}}
\usepackage{balance}
\begin{document}
\title{DIMA: DIffusing Motion Artifacts for unsupervised correction in brain MRI images}
\author{Paolo Angella, Luca Balbi, Fabrizio Ferrando, Paolo Traverso, Rosario Varriale, Vito Paolo Pastore\textsuperscript{*}, Matteo Santacesaria\textsuperscript{*}
\thanks{P. Angella and M. Santacesaria are with MaLGa Center, DIMA, University of Genoa, Italy (e-mail: paolo.angella@edu.unige.it; matteo.santacesaria@unige.it)}
\thanks{V.P. Pastore is with MaLGa Center, DIBRIS, University of Genoa, Italy (e-mail: vito.paolo.pastore@unige.it)}
\thanks{L. Balbi, F. Ferrando, P. Traverso and R. Varriale are with Esaote S.p.A., Italy (e-mail: luca.balbi@esaote.com; fabrizio.ferrando@esaote.com; paolo.traverso@esaote.com; rosario.varriale@esaote.com)}
\thanks{*These authors contributed equally to this work.}
}

\maketitle
\begin{abstract}
Motion artifacts remain a significant challenge in Magnetic Resonance Imaging (MRI), compromising diagnostic quality and potentially leading to misdiagnosis or repeated scans. Existing deep learning approaches for motion artifact correction typically require paired motion-free and motion-affected images for training, which are rarely available in clinical settings. To overcome this requirement, we present DIMA (DIffusing Motion Artifacts), a novel framework that leverages diffusion models to enable unsupervised motion artifact correction in brain MRI. Our two-phase approach first trains a diffusion model on unpaired motion-affected images to learn the distribution of motion artifacts. This model then generates realistic motion artifacts on clean images, creating paired datasets suitable for supervised training of correction networks. Unlike existing methods, DIMA operates without requiring k-space manipulation or detailed knowledge of MRI sequence parameters, making it adaptable across different scanning protocols and hardware. Comprehensive evaluations across multiple datasets and anatomical planes demonstrate that our method achieves comparable performance to state-of-the-art supervised approaches while offering superior generalizability to real clinical data. DIMA represents a significant advancement in making motion artifact correction more accessible for routine clinical use, potentially reducing the need for repeat scans and improving diagnostic accuracy.
\end{abstract}


\section{Introduction}
\IEEEPARstart{M}{\lowercase{agnetic}} Resonance Imaging (MRI) \cite{review1,review2} is a powerful medical imaging technique. However, the technique suffers from a critical limitation: patients must remain extraordinarily still during imaging.
This immobility requirement creates significant challenges across different patient groups. Young children, claustrophobic individuals, and elderly patients often struggle to remain motionless for extended periods. Additionally, some movements are simply unavoidable, such as those caused by breathing and heart function.
The motion-induced artifacts in MRI are uniquely complex, stemming directly from the image formation process. Unlike traditional photography, where movement creates localized blurring, MRI's image construction involves sampling points in Fourier space (known as k-space). In this domain, even minor local movements can dramatically distort the entire image \cite{zaitsev2015motion}. This occurs because MRI images are created by sampling points from the Fourier transform of the object and then computing an inverse Fourier transform with incomplete data. As an inverse problem, these motion-induced errors can be substantially amplified.
To mitigate these challenges, researchers have developed two primary intervention strategies: prospective and retrospective techniques. Prospective methods are implemented during the scanning process, such as collecting redundant data points at different times to compensate for patient movement. However, these approaches are often machine-specific, technique-dependent, and can increase overall scan duration. Retrospective techniques, in contrast, focus on correcting artifacts after the image has been captured. Our proposed method belongs to this second category.
\subsection{Motion artifacts}
While MRI offers superior soft tissue contrast and non-invasive imaging capabilities without ionazing radiations, its relatively long acquisition times make it susceptible to patient movement, which can result in various image artifacts including ghosting, blurring, geometric distortion, and reduced signal-to-noise ratio (SNR). These artifacts can compromise diagnostic accuracy and automated analysis, potentially leading to false positive or negative findings, and may necessitate scan repetition, which increases costs and delays treatment. Motion correction approaches can be classified by several factors: motion type (rigid translation/rotation vs. deformable), timing (inter-image, inter-scan, or intra-scan), pattern (periodic, quasi-periodic, continuous, or sporadic), and in 2D imaging, whether the motion occurs in-plane or through-plane. While faster imaging techniques have reduced acquisition times from dozens of minutes to seconds, these improvements often involve trade-offs in resolution, contrast, or SNR. The ongoing development of motion correction techniques reflects the critical importance of addressing this fundamental challenge in MRI technology, as improving motion correction capabilities directly impacts clinical outcomes and research validity \cite{godenschweger2016motion}.
\subsection{Classical techniques}
Classical methods to mitigate motion artifacts in MRI primarily focus on reducing patient movement and correcting the acquired data. Physical restraints and breath-holding techniques are employed to minimize motion during image acquisition. Prospective motion correction \cite{maclaren2013prospective} actively adjusts scan parameters in real-time to compensate for detected movement, while retrospective methods apply post-processing algorithms to realign and correct motion-affected image segments. Additionally, navigator echoes, temporal filtering, and multi-shot acquisition strategies with motion-robust reconstruction algorithms \cite{zaitsev2015motion} can help reduce the impact of motion on image quality. These techniques, while effective, often require careful optimization and may not fully address complex motion patterns, particularly in challenging clinical scenarios.
\subsection{Deep learning}
Recent advances in deep learning have enabled significant progress in retrospective motion artifact correction for MRI \cite{review1,review2}. These methods rely on training machine learning models, predominantly deep neural networks, through supervised approaches where data quality is paramount. Two primary strategies exist for obtaining suitable training data. The first involves acquiring paired scans of the same subject—one motion-free and one with motion artifacts. However, this approach presents substantial limitations: such paired acquisitions rarely occur in clinical practice, requiring expensive dedicated experiments that may not accurately represent involuntary patient movements encountered in real-world scenarios. The second approach involves simulating motion artifacts through k-space manipulation, which, while eliminating the need for additional scans, requires detailed knowledge of the MRI acquisition parameters and may produce artifacts that inadequately represent clinical reality. Our proposed method addresses these limitations by utilizing unpaired motion-affected and motion-free images from routine clinical examinations, eliminating the requirement for k-space modification or sequence-specific technical information, thereby enabling broader application across various scanning protocols and hardware configurations.
\subsection{Literature review}
Prospective methods either use sampling patterns that sample the same points in the k-space at different times to estimate and correct the patient motion \cite{propeller, liao1997reduction} or use external tracking devices such as cameras to achieve the same objective \cite{todd2015prospective, white2010promo}.
Among the retrospective methods we cite the ones using compressed sensing techniques \cite{usman2013motion,6556636} that leverage a priori knowledge on MR images to perform the motion correction task.
Numerous deep learning approaches have been proposed for MRI motion artifact correction. In \cite{duffy2018retrospective}, CNNs were used for motion correction in 3D MRI data, while \cite{johnson2019conditional} employed a conditional generative adversarial network (cGAN) for artifact reduction. \cite{liu2020motion} introduced a deep residual network with densely connected multi-resolution blocks, and \cite{al2022stacked} developed stacked U-Nets with self-assisted priors, such as contiguous slices, for improved corrections. In \cite{kuzmina2022autofocusing}, a neural network-based regularization term was integrated with Autofocusing, a classic optimization-based method, and \cite{levac2023accelerated} employed a diffusion model for motion correction. Recently, \cite{oh2023annealed} proposed an annealed score-based diffusion model, which consists in iteratively applying forward and reverse diffusion processes. All of these methods rely on supervised learning and require paired motion-corrupted and motion-free images, which are often unavailable and necessitate the simulation of motion artifacts. 
\\
Different approaches \cite{liu2021learning,ghodrati2021retrospective,oh2021unpaired,armanious2020unsupervised,kim2025unsupervised}  faced the data availability problem by using an image-to-image translation framework, exploiting general techniques such as \cite{isola2017image}, and with a framework inspired by  \cite{armanious2019unsupervised}. 
Exploiting synthetic data, these methods enabled artifact removal without paired training data through innovative approaches like adversarial autoencoders, k-space subsampling, and novel reconstruction methods, thereby enhancing image quality across respiratory, brain, and dynamic imaging scenarios. 
Specifically, in \cite{liu2021learning}, the authors employ an adversarial network to learn the translation mappings between the motion-affected and motion-free images; in \cite{ghodrati2021retrospective} instead they trained an adversarial autoencoder network in a way that  forces the encoder to remove motion artifacts; in \cite{oh2021unpaired} they developed an unsupervised deep learning scheme through outlier rejecting bootstrap sub-sampling and aggregation, using data in the k-space;  \cite{armanious2020unsupervised} introduced an adversarial framework with a new generator architecture and loss function to perform unsupervised correction; finally in \cite{kim2025unsupervised} they improved the vanilla CycleGAN by adding loss regularization with a module that estimates the motion severity of the input image.

All these works were originally evaluated on synthetic or proprietary datasets. However, a subsequent study \cite{safari23} provided a comparative evaluation of the aforementioned techniques using the MR-ART dataset, which we also use in this work. The work \cite{safari23} also introduced the first application of diffusion models for motion artifact correction, where motion-free images are generated directly from their motion-affected counterparts. Their diffusion-based method significantly outperformed state-of-the-art approaches based on image-to-image translation. Recently, in \cite{angella2025} the authors performed a critical evaluation of the usage of diffusion models for the motion-artifact correction task, showing correlations between scan plane and correction quality and especially, how harmful hallucinations can emerge. 
Similarly to previous works, we exploited diffusion models for motion artifact correction; however, differently from \cite{angella2025, safari23}, rather than using the diffusion model for generating motion-free images, we exploit it to generate paired motion-affected motion-free images. Our generated couples are further used in a hybrid framework, where a U-net is trained to predict motion-free images. 
\section{Methods}
\begin{figure}
\centering
\includegraphics[width=0.45\textwidth]{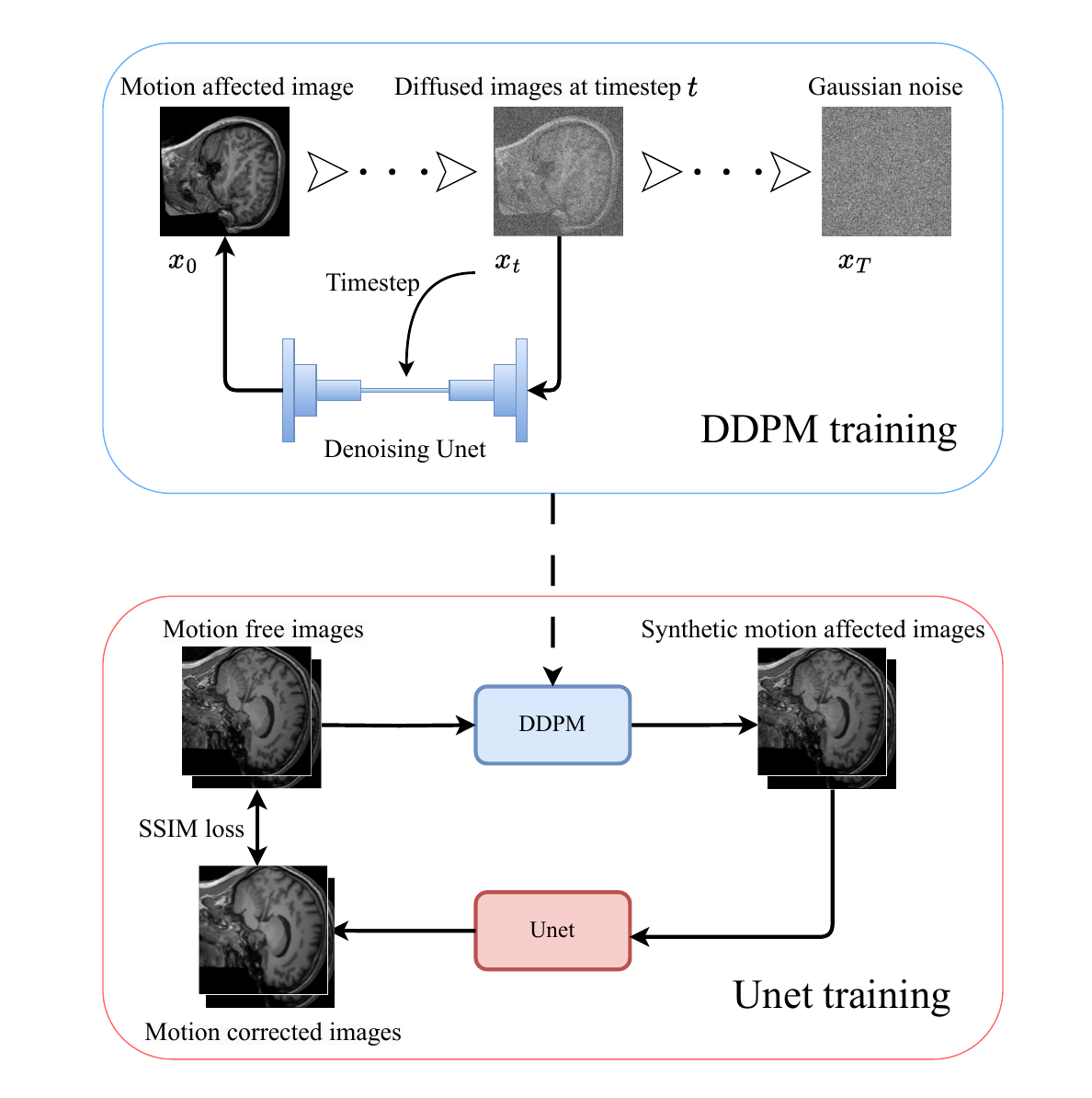}
\caption{Our implementation follows a two-phase approach. In the first phase, a diffusion model (DDPM) is trained on a dataset of motion artifact-affected images. In the second phase, the trained model is used to introduce motion artifacts into clean images, generating paired datasets. These pairs enable supervised training in the subsequent step. See Algorithm \ref{alg:motion_augmentation} for more details.}
\label{All}
\end{figure}
Our proposed method generates paired datasets for supervised training using initially unpaired data. The process involves two separate datasets: one containing images affected by motion artifacts and another with motion-free images. These datasets may originate from different patients.

Our method is depicted in Figure \ref{All}. First, we train a generative model, specifically a Diffusion model \cite{NEURIPS2020_4c5bcfec}, on the motion-affected images. This model learns the distribution of motion artifacts and can generate similar images. Instead of using the model solely for image generation, we condition it on the motion-free images, allowing us to produce motion-affected versions of them. This results in a new dataset where each motion-free image is paired with a corresponding motion-affected version.

With this paired dataset, we train a supervised model to correct motion artifacts. The performance of this model is then evaluated by comparing it to the same model trained using alternative data simulation techniques, as well as with real pairs of motion-free and motion-affected images when available.
\subsection{Denoising diffusion probabilistic model}
Denoising Diffusion Probabilistic Models (DDPMs), commonly referred to as Diffusion Models for brevity, are generative models that create data by reversing a diffusion process. In the forward diffusion process, Gaussian noise is progressively added to an initial data point $\mathbf{x}_0$ (e.g., an MRI image) over $T$ timesteps, resulting in a sequence of latent variables $\mathbf{x}_1, \dots, \mathbf{x}_T$. Each transition is governed by the following rule:
\begin{equation}
    q(\mathbf{x}_t | \mathbf{x}_{t-1}) = \mathcal{N}(\mathbf{x}_t; \sqrt{\alpha_t} \mathbf{x}_{t-1}, (1-\alpha_t) \mathbf{I}),
\end{equation}
where $\alpha_t$ are hyperparameters of the variance schedule parameters. The reverse process, that is learned with a neural network, learns to denoise each $\mathbf{x}_t$ to recover $\mathbf{x}_{t-1}$:
\begin{equation}
    p_\theta(\mathbf{x}_{t-1} | \mathbf{x}_t) = \mathcal{N}(\mathbf{x}_{t-1}; \mu_\theta(\mathbf{x}_t, t), \Sigma(\mathbf{x}_t, t)).
\end{equation}
This architecture enables DDPMs to generate samples by progressive denoising starting from pure Gaussian noise.
In this work we want to leverage the DDPM to introduce motion artifacts, as it typically generates images by starting from random noise and performing $T$ denoising steps. In our approach, described in Alg. \ref{alg:motion_augmentation}, we adapt the DDPM to begin with a motion-free image, add Gaussian noise progressively up to a certain step $n$ (where $n < T$), and then allow the model to denoise for the remaining $n$ steps. This method enables the model to introduce realistic motion artifacts by simulating the forward diffusion process up to step $n$ followed by partial denoising that retains these characteristics while introducing the artifacts. The choice of $n$ is critical to achieving a balance; if $n$ is too large, the added artifacts may appear exaggerated or unrealistic, diverging from the characteristics of motion artifacts in real-world data. Conversely, if $n$ is too low, the artifacts may not be prominent enough, making them ineffective for generating meaningful paired training data. By tuning $n$, we ensure that the generated motion-affected images are both realistic and diverse, enabling the creation of high-quality paired datasets for supervised training.

\begin{algorithm}
\caption{Diffusion-based motion simulation. The algorithm used to add simulated motion artifacts to motion-free images; this is done after training the network $\epsilon_0$ on the motion-affected images}
\label{alg:motion_augmentation}
\begin{algorithmic}[1]
\State \textbf{input} $\mathbf{Y},n, \left\{ \alpha_t, \bar \alpha_t, \sigma_t \right\}_{t=1}^{n}, \mathbf{\epsilon}_\theta$
\State $\mathbf{z} \sim \mathcal{N}(0, \mathbf{I})$
\State $\mathbf{x}_n =  \sqrt{\bar{\alpha}_n} \mathbf{Y} + (1 - \bar{\alpha}_n) \mathbf{z}$

\For{$t = n, \dots, 1$}
\State if {$t > 1$} $\mathbf{z} \sim \mathcal{N}(0, \mathbf{I})$, else $\mathbf{z} = 0$
    \State $\mathbf{x}_{t-1} = \frac{1}{\sqrt{\alpha_t}} \left( \mathbf{x}_t - \frac{1 - \alpha_t}{\sqrt{1 - \bar{\alpha}_t}} \mathbf{\epsilon}_\theta (\mathbf{x}_t, t) \right) + \sigma_t \mathbf{z}$
\EndFor
\State \textbf{return} $\mathbf{x}_0$
\end{algorithmic}
\end{algorithm}
\subsection{U-net}
Following the generation of simulated data, we employ a U-Net architecture \cite{ronneberger2015u} to learn a mapping between motion-corrupted images and their artifact-free counterparts. The U-Net’s design is particularly well-suited for biomedical image analysis due to its encoder-decoder structure with skip connections, which preserves fine spatial details while capturing hierarchical feature representations.
To train the network, we utilize paired datasets where motion-affected images (simulated via our framework) serve as inputs, and their corresponding ground-truth, artifact-free images act as targets. During training, we optimize using the SSIM loss function.
Once trained, the U-Net operates as a standalone artifact correction model, processing unseen motion-degraded images in inference. To validate performance, we quantify restoration accuracy using as metrics  structural similarity index (SSIM), mean squared error (MSE) and peak signal-to-noise ratio (PSNR).
\section{Experiments}
\begin{figure}
\centering
\includegraphics[width=0.45\textwidth]{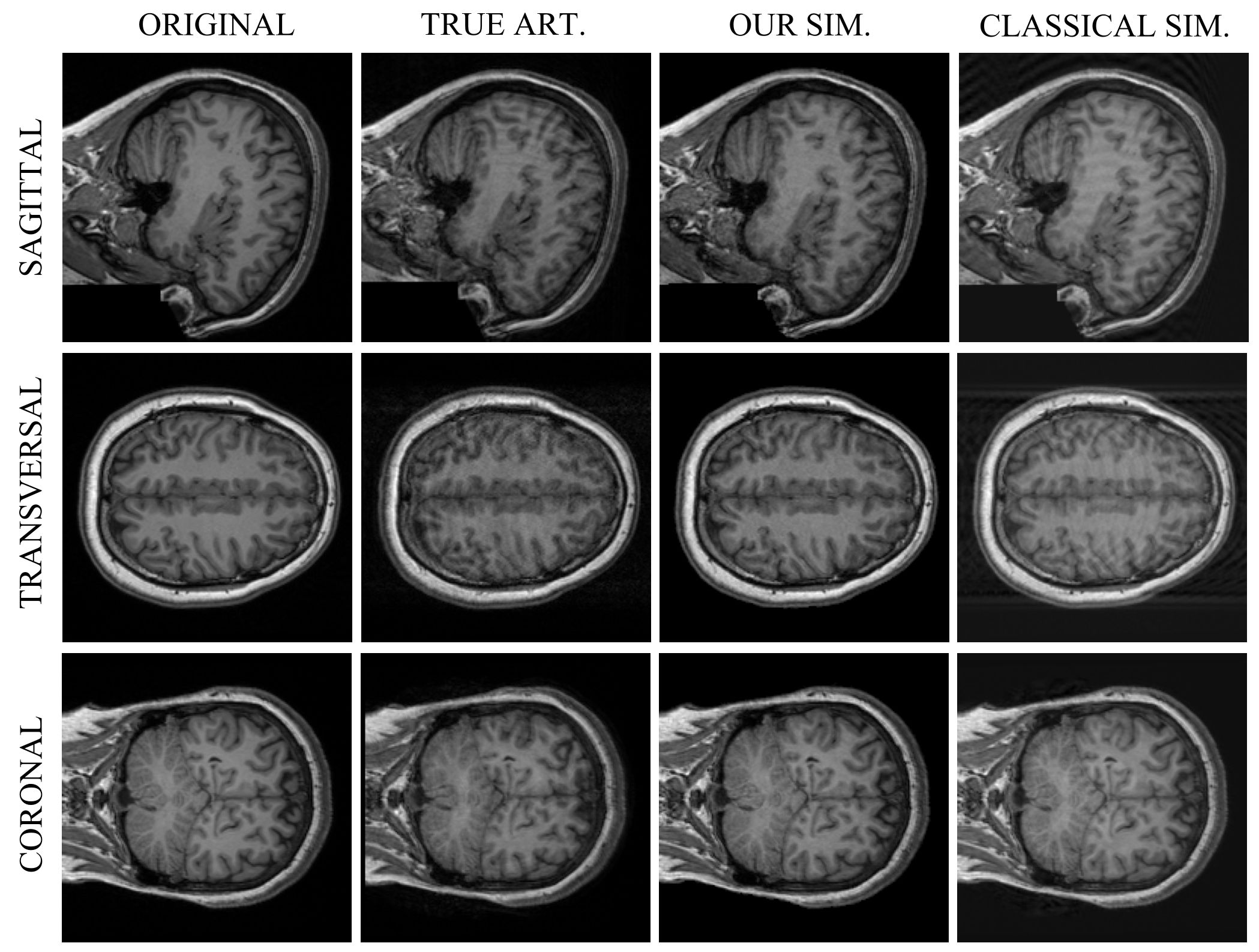}
\caption{Example of the the two motion simulation techniques used next to the original images. \label{noisy}}
\end{figure}
\subsection{Dataset}
Our experiments were conducted using two datasets. The first dataset is the MR-ART dataset \cite{MR-ART}, which includes brain MRI scans from 144 patients, where we extracted around 100 images from each scan for each possible view (sagittal, transversal and coronal). For each patient, there are three scans: one motion-free scan acquired while the patient remained still and two motion-affected scans acquired while the patient purposefully moved their head to induce artifacts. Additionally, we utilized simulated motion artifact data provided in \cite{olsson2024simulating}, which generated four motion-affected simulations for each motion-free scan in the MR-ART dataset. This is done through k-space alteration techniques that mathematically model the complex physical processes underlying
image degradation during magnetic resonance scanning, we point out that generating this simulations requires to know the details of the particular MRI sequences used to acquire the scans.
The second dataset \cite{in-s}, which we refer to as IN-SCANNER, consists of MRI scans from 25 patients, each with one motion-free and one motion-affected scan.
It is important to note that paired motion-free and motion-affected MRI scans are exceptionally rare, particularly for non-brain MRI. This scarcity arises because accidental patient movement during scans is challenging to capture in a controlled manner. The motion artifacts in the MR-ART dataset were generated through purposeful movement, which differs from the unintentional and often irregular motion seen in real-world clinical settings.
Both datasets comprise 3D volumes, from which we extracted slices from all possible acquisition planes (sagittal, transversal and coronal) for our experiments. As part of the preprocessing pipeline, we applied normalization and performed registration to align the motion-free and motion-affected images in the MR-ART dataset, as the volumes were often misaligned.
\subsection{Implementation details}
\subsubsection{\textbf{Diffusion model}}

We trained a Denoising Diffusion Probabilistic Model (DDPM) based on the implementation from \cite{ddpm-github}. The model consists of a U-net architecture that accepts both an image and an integer representing the denoising time-step as inputs. We trained this DDPM on motion-affected images from 30 patients in the dataset (for reference, there are around 120 images for each patient), using an additional 15 for validation. 500 time-steps were used and we implemented early stopping to prevent overfitting. These patients' data were excluded from all subsequent experiments to maintain data independence.
To generate synthetic motion-affected images starting from the available motion-free ones, and employing the trained diffusion model, we select an integer $n$ smaller than the total number of time-steps, add noise to each image slice according to the DDPM scheduler's specifications, and feed the noisy image to be denoised at the $n$-th timestep (see Algorithm \ref{alg:motion_augmentation}).
 This process can be repeated multiple times to improve the result by simply taking the image generated and feeding it back to the DDPM, repeating the process. This approach is used because running multiple iterations allows for the introduction of more artifacts while minimizing hallucinations, compared to a single, deeper denoising process with more steps.
There's a crucial balance to strike when choosing the timestep value. A timestep that's too large results in heavily hallucinated images, while one that's too small produces images too similar to the originals. Either extreme yields pairs that are less useful for training. See Figure \ref{noisy} for a qualitative comparison of our diffusion-based motion artifact simulation with real motion artifacts from the MR-ART dataset and synthetic motion artifact from \cite{olsson2024simulating}.
\subsubsection{\textbf{U-net}}
For rigorous comparative analysis, we simultaneously trained identical U-net neural network architectures using three alternative image sets. The first alternative training set is comprised of ground truth images sourced directly from the MR-ART dataset, providing a baseline of what can be obtained with paired data. The second alternative training set consisted of images with simulated motion artifacts generated by \cite{olsson2024simulating}. Finally we have our own proposed method that uses synthetic images created by the diffusion model 
trained on real motion-affected images. 
In all cases we use the original motion-free images paired with the motion-affected real/simulated ones for training.
By employing diverse image sets to train these networks, we developed a solid experimental framework that allows us to thoroughly assess the effectiveness and adaptability of our generation method. This methodical comparison allows us to assess not only the efficacy of our proposed method, but also its potential advantages over traditional motion artifact correction techniques.
Our experimental design intentionally maintained consistent network architectures across all training scenarios, ensuring that any observed performance differences could be attributed directly to variations in the training data rather than network structure. This approach provides a motion-free, controlled methodology for understanding the nuanced challenges of motion artifact correction in medical imaging.
\subsubsection{\textbf{Data and parameter selection}}
\begin{figure}
\centering
\includegraphics[width=0.45\textwidth]{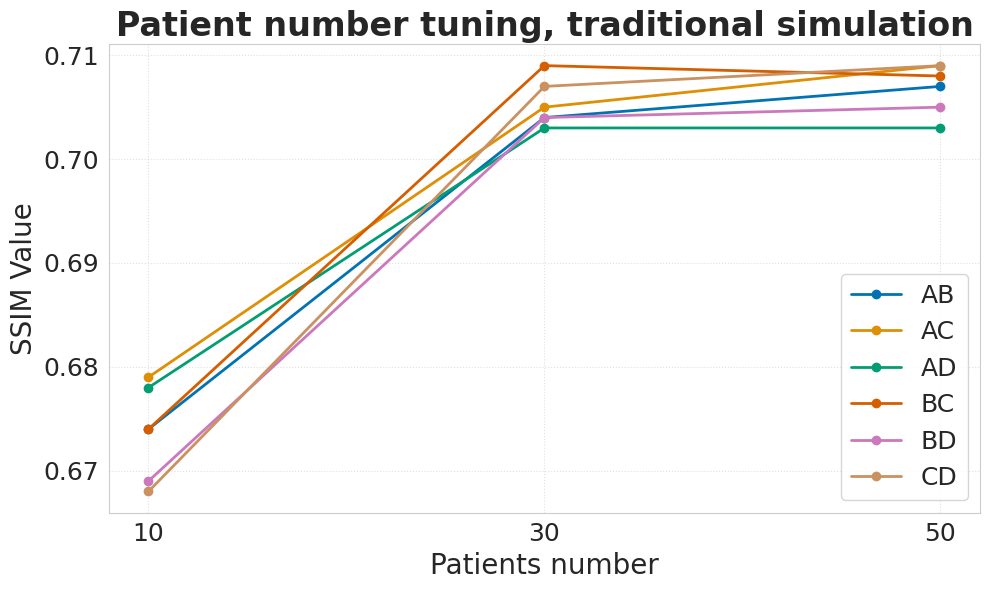}
\caption{Graph comparing the performance obtained using different simulation parameters from \cite{olsson2024simulating} with varying numbers of patients.}
\label{graph_a}
\end{figure}

All U-net networks were trained with a learning rate of 0.001, a batch size of 6, and the Structural Similarity Index Measure (SSIM) \cite{ssim} as the loss function. We implemented early stopping to prevent overfitting. The test set consisted of ground truth pairs from the MR-ART dataset, and performance was evaluated using three metrics: Structural Similarity Index Measure (SSIM), Normalized Mean Squared Error (NMSE), and Peak Signal-to-Noise Ratio (PSNR).
The MR-ART dataset pairs each motion-free reference image with two motion-affected variants. To maintain consistency in our U-Net training, we adopted the same approach by using two simulated motion-affected images per motion-free image in both simulated methods. Since the dataset provided by \cite{olsson2024simulating} contains four distinct sets of motion-affected images (A, B, C, and D) for each reference image, we conducted separate U-Net training experiments for every possible combination of these sets (results in Table \ref{tab:ABCD}). This systematic approach enabled us to identify the optimal image set for training.
In addition to image sets, we investigated the number of patients required to achieve optimal performance, as motion simulations are computationally expensive. We trained all U-Net configurations using 10, 30, and 50 patients while maintaining a fixed validation set of 15 patients. These tests were performed exclusively on sagittal slices, with the selected parameters subsequently applied to the other two image views. During training, validation was performed using the same input images as those in the training set. However, to more accurately represent real-world performance, the data selection results (SSIM metrics shown in Figure \ref{graph_a}) were calculated using original paired images from the MR-ART dataset. The results show that 30 patients represents a sufficient sample size, showing significant improvement over 10 patients while performing comparably to 50 patients. Detailed metrics for U-Nets trained with 30 patients across different training sets are presented in Table \ref{tab:ABCD}.
\begin{table}
\begin{center}
\begin{tabular}{llll}
\hline
Sets used & SSIM & NMSE & PSNR \\
\hline
AB & $0.704 \pm 0.067$ & $0.117 \pm 0.06$ & $21.3 \pm 2.0$ \\
AC & $0.705 \pm 0.069$ & $0.113 \pm 0.071$ & $21.3 \pm 2.2$ \\
AD & $0.703 \pm 0.070$ & $0.114 \pm 0.064$ & $21.3 \pm 2.2$ \\
BC & $\textbf{0.709} \pm 0.068$ & $\textbf{0.112} \pm 0.063$ & $\textbf{21.4} \pm 2.1$ \\
BD & $0.704 \pm 0.068$ & $\textbf{0.112} \pm 0.059$ & $21.3 \pm 2.1$ \\
CD & $0.707 \pm 0.069$ & $0.117 \pm 0.063$ & $21.1 \pm 2.1$ \\
\hline
\end{tabular}
\vspace{6pt} 
\caption{Performance metrics for different dataset combinations of the data made available by \cite{olsson2024simulating}.}
\label{tab:ABCD}
\end{center}
\end{table}
Notably, all combinations yielded highly similar results. We ultimately selected combination "BC" due to its marginally higher SSIM and lower NMSE values, though we emphasize that no significant differences exist between combinations.

\begin{figure}
\centering
\includegraphics[width=0.45\textwidth]{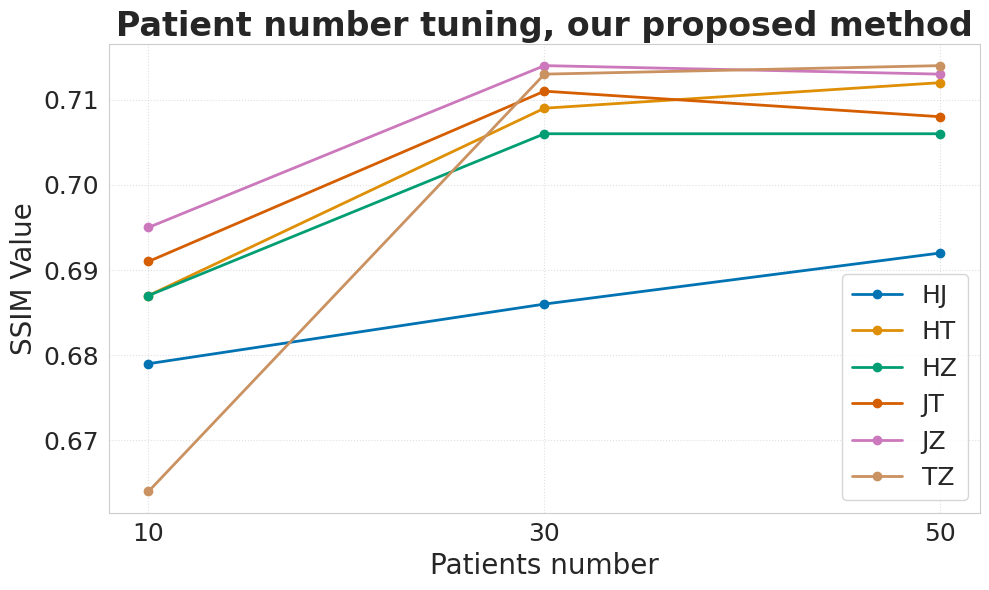}
\caption{Graph comparing the performance obtained using different simulation parameters in our proposed methods with varying numbers of patients.}
\label{graph_h}
\end{figure}

For consistency, we also evaluated four potential data simulation parameters (H, J, T, and Z) against the control set and those from \cite{olsson2024simulating}. The parameters for each set, shown in Table \ref{tab:TZHJ}, primarily involved the number of denoising steps, which determines how far deep in the DDPM architecture the process starts, and iterations, meaning how many times Algorithm \ref{alg:motion_augmentation} is repeated. These values were heuristically selected to balance motion artifacts and hallucinations. Following the same protocol, we tested all combinations with 10, 30, and 50 patients (results in Figure \ref{graph_h}), again identifying 30 patients as sufficient. Detailed metrics for these U-Nets appear in Table \ref{tab:Sel-htz}. While results showed smaller variations between parameter choices, we selected combination JZ due to its slight performance advantage. We emphasize that the patients used for parameter selection were distinct from those in the final test set.
\newline
\begin{table}
\begin{center}
\begin{tabular}{lll}
\hline
Set & Denoising steps & Iterations \\ \hline
T & $120$ & $4$ \\ 
Z & $170$ & $5$ \\ 
H & $280$ & $2$ \\ 
J & $330$ & $1$ \\ \hline
\end{tabular}
\vspace{6pt} 
\caption{Parameters of motion simulation for the diffusion model.}
\label{tab:TZHJ}
\end{center}
\end{table}
\begin{table}
\centering
\begin{tabular}{llll}
\hline
Label & SSIM & NMSE & PSNR \\
\hline
HJ & $0.686 \pm 0.068$ & $0.122 \pm 0.061$ & $19.5 \pm 2.0$ \\
HT & $0.709 \pm 0.077$ & $0.106 \pm 0.058$ & $\textbf{20.8} \pm 2.3$ \\
HZ & $0.706 \pm 0.074$ & $0.11 \pm 0.058$ & $20.1 \pm 2.1$ \\
JT & $0.711 \pm 0.071$ & $0.109 \pm 0.059$ & $20.3 \pm 2.1$ \\
JZ & $\textbf{0.714} \pm 0.068$ & $0.111 \pm 0.057$ & $20.0 \pm 2.0$ \\
TZ & $0.713 \pm 0.068$ & $\textbf{0.106} \pm 0.057$ & $20.5 \pm 2.1$ \\
\hline
\end{tabular}
\vspace{6pt} 
\caption{Performance metrics for different dataset combinations of our simulated images.}
\label{tab:Sel-htz}
\end{table}
\subsection{Results}
We compare the performance of different U-Net models trained under varying conditions. To facilitate this analysis, we refer to the following methods:
\begin{enumerate}
    \item \textbf{SIMUNET}: A U-Net trained with traditional motion simulation techniques, utilizing data from \cite{olsson2024simulating}.
    \item \textbf{DIMA}: A U-Net employing our proposed diffusion-based motion simulation technique.
    \item \textbf{SUPUNET}: A U-Net trained in a supervised manner using pairs of real motion-affected and motion-free images, serving as an upper-bound benchmark for the performance of the two preceding methods.
\end{enumerate}

\begin{figure}
\centering
\includegraphics[width=0.49\textwidth]{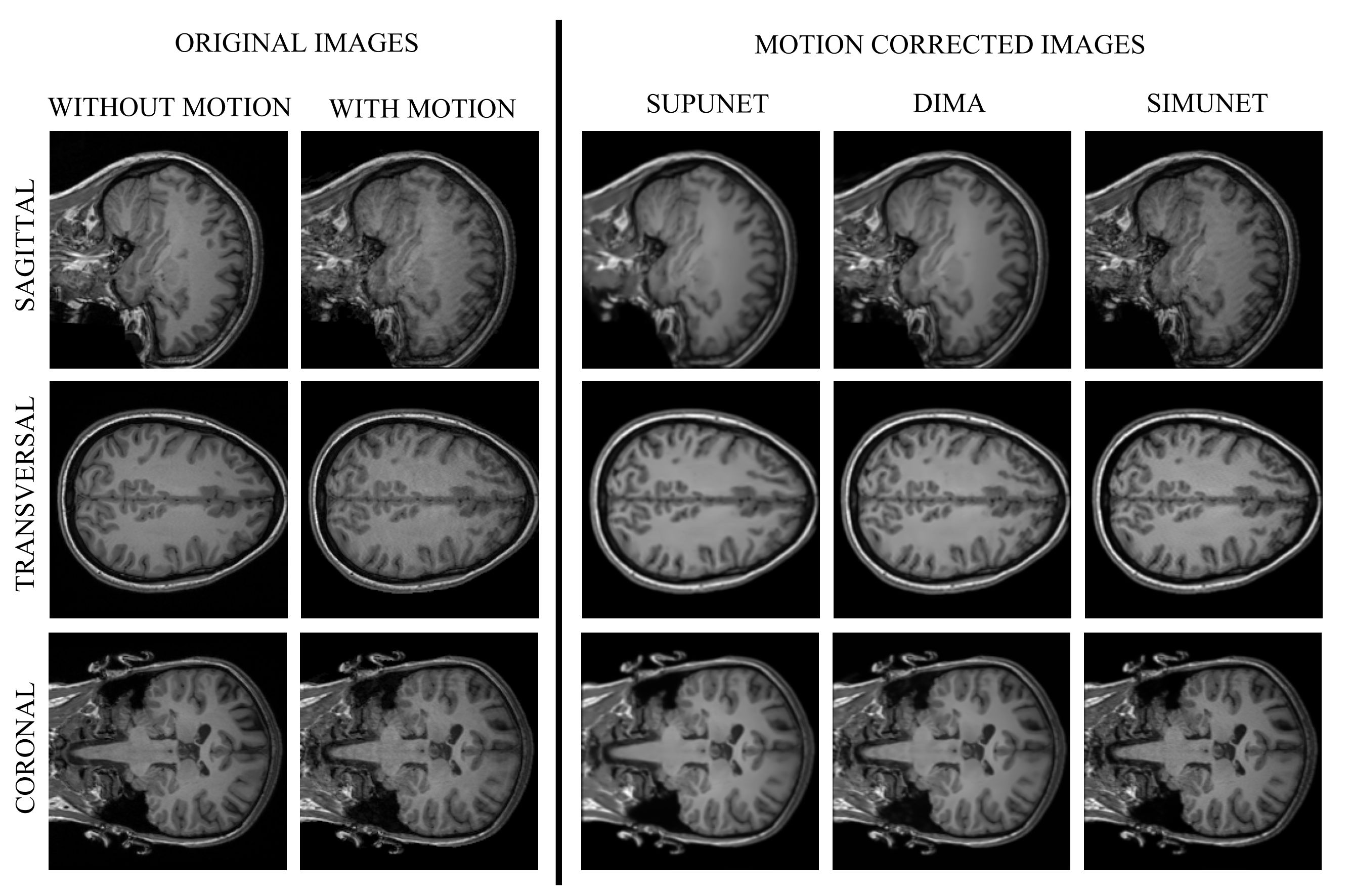}
\caption{Example of the output of the various U-nets on the test set. From left to right: ground truth motion-free and motion-affected images, three different outputs of the three different networks. \label{confronto}}
\end{figure}

\begin{table}
\centering
\begin{tabular}{llll}
\hline
Label & SSIM & NMSE & PSNR \\
\hline
SIMUNET \cite{olsson2024simulating} & $0.786 \pm 0.035$ & $0.064 \pm 0.016$ & $\textbf{23.6} \pm 1.0$ \\
DIMA & $\textbf{0.788} \pm 0.036$ & $\textbf{0.058} \pm 0.017$ & $23.0 \pm 1.2$ \\
\hline
SUPUNET & $0.820 \pm 0.030$ & $0.052 \pm 0.015$ & $23.5 \pm 1.3$ \\
\hline
CycleGAN \cite{safari23} & $0.670$ & $0.225 $ & $21.7$ \\
Pix2Pix \cite{safari23} & $0.640$ & $0.242$ & $21.4$ \\
Unet \cite{safari23} & $0.730$ & $0.195 $ & $22.5 $ \\
Diff \cite{safari23} & $0.680$ & $0.191 $ & $21.9 $ \\
\hline
\end{tabular}
\vspace{6pt} 
\caption{Experiments results for the sagittal acquisition plane.}
\label{tab:MR-sagg}
\end{table}
The comparative performance across different anatomical views and datasets reveals several key insights. See Figure \ref{confronto} for a qualitative comparison. For sagittal views (Table \ref{tab:MR-sagg}), our approach demonstrates comparable performance with the ground truth benchmark (SSIM: $0.788 \pm 0.036$ vs $0.820 \pm 0.030$), while substantially outperforming traditional architectures like CycleGAN and Pix2Pix.
The view-specific analysis reveals interesting variations in performance. While our method shows results in line to the simulation-based approach in all the views (SSIM: $0.852 \pm 0.037$ vs $0.858 \pm 0.034$, Table \ref{tab:MR-tran}, and SSIM: $0.807 \pm 0.057$ vs $0.806 \pm 0.040$, Table \ref{tab:MR-coronal}), it performs better compared to ground truth benchmark in the transversal view. This anatomical dependency likely reflects differences in motion artifact complexity across imaging planes, with transversal views potentially containing more challenging rotational artifacts that our current simulation parameters capture less effectively.
Crucially, when tested on the completely independent IN-SCANNER dataset (Table \ref{tab:altro}), for which we used 11 patients for the training of the diffusion model, 11 for the training of the U-nets and 4 in test, our approach demonstrates to generalize well, with SSIM ($0.743 \pm 0.076$) and PSNR ($23.2 \pm 1.7$ dB) metrics closely matching performance obtained when training with real pairs (SSIM: $0.761 \pm 0.065$, PSNR: $22.8 \pm 1.5$ dB). This is particularly notable given that our method requires no k-space manipulation or sequence-specific knowledge - a significant practical advantage over traditional acquisition-dependent approaches. 
\begin{table}
\centering
\begin{tabular}{llll}
\hline
Label & SSIM & NMSE & PSNR \\
\hline
SIMUNET \cite{olsson2024simulating} & $\textbf{0.858} \pm 0.034$ & $\textbf{0.046} \pm 0.019$ & $\textbf{23.7} \pm 1.5$ \\
DIMA & $0.852 \pm 0.037$ & $0.047 \pm 0.019$ & $23.1 \pm 1.5$ \\
\hline
SUPUNET &$0.863 \pm 0.029$ & $0.054 \pm 0.015$ & $22.0 \pm 1.2$ \\
\hline
\end{tabular}
\vspace{6pt} 
\caption{Experiments results for the transversal acquisition plane.}
\label{tab:MR-tran}
\end{table}

\begin{table}
\centering
\begin{tabular}{llll}
\hline
Label & SSIM & NMSE & PSNR \\
\hline
SIMUNET \cite{olsson2024simulating} & $0.806 \pm 0.040$ & $0.059 \pm 0.020$ & $\textbf{23.6} \pm 1.3$ \\
DIMA & $\textbf{0.807} \pm 0.057$ & $\textbf{0.053} \pm 0.020$  & $23.4\pm 1.4$ \\
\hline
SUPUNET &$0.832  \pm 0.034$ & $0.055 \pm 0.017$ & $22.8 \pm 1.3$ \\
\hline
\end{tabular}
\vspace{6pt} 
\caption{Experiments results for the coronal acquisition plane.}
\label{tab:MR-coronal}
\end{table}

\begin{table}[!t]
\centering
\begin{tabular}{llll}
\hline
Label & SSIM & NMSE & PSNR \\
\hline
DIMA & $0.743 \pm 0.076$ & $0.155 \pm 0.054$  & $23.2\pm 1.7$ \\
\hline
SUPUNET & $0.761 \pm 0.065$ & $0.156 \pm 0.046$ & $22.8 \pm 1.5$ \\
\hline
\end{tabular}
\vspace{6pt} 
\caption{Experiments results for the alternative IN-SCANNER dataset, for the sagittal acquisition plane.}
\label{tab:altro}
\end{table}

\section{Conclusions}
In this work, we proposed a novel technique that leverages diffusion models to address the challenge of requiring paired data for machine learning training in MRI motion artifact correction. By utilizing diffusion models, our approach generates high-quality synthetic paired data, effectively circumventing the need for real paired datasets, which are often difficult to obtain in medical imaging. Three key findings emerge:
\begin{enumerate}
    \item Our simulation framework achieves performance comparable to state-of-the-art motion artifact simulation while eliminating the need for k-space manipulation
\item View-specific performance variations suggest opportunities for plane-adaptive parameter tuning
\item The strong cross-dataset generalization indicate clinical applicability beyond controlled simulation environments
\end{enumerate}
The marginal performance differences between methods (typically less than 3\% SSIM variation) suggest that the choice between approaches may ultimately depend on clinical constraints rather than pure metric superiority. Our method provides a viable alternative for scenarios requiring black-box acquisition compatibility.
Importantly, our method achieved these results without requiring k-space sampling information or detailed knowledge of the MR sequence parameters, representing a significant practical advantage over simulation-based approaches. The consistent performance across different anatomical views, combined with competitive metrics compared to both simulated and real reference data, demonstrates the robustness of our approach for clinical applications. However, challenges such as computational cost and the need to simulate more complex motion patterns remain areas for improvement. Furthermore, this framework could be extended to tackle more general blind inverse problems, such as image deblurring or super-resolution, where paired data is scarce and underlying degradation processes are unknown. These results encourage further exploration of diffusion-based methods not only in MRI but also across broader imaging domains, potentially revolutionizing unsupervised correction techniques.
\section{Compliance with Ethical Standards}
This study was conducted retrospectively using ethically acquired publicly
available human subject data. The authors have no interests to disclose.
\section{Acknowledgments}
\label{sec:acknowledgments}
This material is supported by the Air Force Office of Scientific Research (a.n. FA8655-23-1-7083). Co-funded by Regional Problem FSE+ (point 1.2 of attach. IX of Reg. (UE) 1060/2021) and European Union - Next Generation EU (FAIR ``Future Partnership Artificial Intelligence Research'', CUP J53C22003010006 to MS; PRIN 2022B32J5C ``Inverse problems in PDE: theoretical and numerical analysis", CUP D53D23005770006 to MS). The research was supported in part by the MIUR Excellence Department Project awarded to Dipartimento di Matematica, Università di Genova, CUP D33C23001110001.

\bibliographystyle{IEEEtran}

\end{document}